\documentclass[a4paper]{emulateapj}
\usepackage{amsmath,amssymb}
\usepackage{graphicx,epsfig,color}
\usepackage{ulem}

\renewcommand{\d}{\mathrm{d}}

\newcommand{\bea}{\begin{eqnarray}}
\newcommand{\eea}{\end{eqnarray}}
\newcommand{\be}{\begin{equation}}
\newcommand{\ee}{\end{equation}}
\newcommand{\rund}[1]{\left(#1\right)}
\newcommand{\eck}[1]{\left[ #1 \right]}

\newcommand{\vc}[1]{\mbox{\boldmath $#1$}}

\newcommand{\msun}{h^{-1}M_{\odot}}

\def\elabel#1{\label{eq:#1}}

\def\oii{{[O \sc ii]}\,}

\begin{document}

\title{Biases in physical parameter estimates through
differential lensing magnification}

\author{Xinzhong Er \altaffilmark{1},
    Junqiang Ge \altaffilmark{1},
    Shude Mao \altaffilmark{1,2}
}%
\email{xer@nao.cas.cn}
\altaffiltext{1}{National Astronomical Observatories, Chinese Academy
  of Sciences, Beijing 100012, China}
\altaffiltext{2}{Jodrell Bank Centre for Astrophysics, University of
  Manchester, Alan Turing Building, Manchester M13 9PL, UK}

\begin{abstract}
We study the lensing magnification effect on background galaxies.
Differential magnification due to different magnifications of
different source regions of a galaxy will change the lensed composite
spectra. The derived properties of the background galaxies are therefore
biased. For simplicity, we model galaxies as a superposition of an
axis-symmetric bulge and a face-on disk in order to study the
differential magnification effect on the composite spectra. We find
that some properties derived from the spectra (e.g., velocity
dispersion, star formation rate and metallicity) are modified.
Depending on the relative positions of the source and the lens, the
inferred results can be either over- or under-estimates of the true
values. In general, for an extended source at strong lensing regions
with high magnifications, the inferred physical parameters
(e.g. metallicity) can be strongly biased. Therefore detailed lens
modelling is necessary to obtain the true properties of the lensed
galaxies.
\end{abstract}
%
%
%

\section{Introduction}

The properties of high redshift objects, mostly galaxies, provide
important information on the early evolution of galaxies and the role
they played in the cosmic reionization
\citep[e.g.][]{2010hdfs.book.....L,2012RAA....12..865F}. Observations
of the high redshift galaxies are challenging due to large distances and
faint magnitudes. With the growing power of telescopes, the number of
high redshift galaxies found is increasing rapidly, either by deep imaging
or the color selection (the $V$-band, $r$-band ``dropout" technique)
\citep[e.g.][]{2004ApJ...607..704S, 2009A&A...498..725H}. Moreover,
the $Y-$band and $J-$band dropouts can potentially detect objects at
redshift $z\sim8$ \citep[e.g.][]{2012ApJ...761..177Y}. Galaxy
clusters, as nature telescopes can enhance the ability to detect
high redshift galaxies \citep{1990Ap&SS.170..283S}. The method has
been implemented to study galaxies over a wide range of redshifts
\citep[e.g.][]{2001ApJ...560L.119E, 2009ApJ...706.1201B,
  2012ApJ...745..155H} as well as Ly$\alpha$ spheres
\citep{2007ApJ...666...45L}. It has been shown that the search for the
high-$z$ galaxies ($z>7$) in a lensing field is more efficient than
that in blank fields, and the maximum efficiency is reached for
lens clusters at $z\sim 0.1-0.3$ \citep{2010A&A...509A.105M}.

Gravitational lensing provides a direct way for studying the mass
distribution of the large scale structures in the universe as well as
galactic- and cluster-sized halos
\citep[e.g.][]{2001PhR...340..291B,2010ARA&A..48...87T}.  On the other
hand, the light from distant galaxies can be magnified by several
orders of magnitude by the gravitational potential well of massive
galaxy clusters. The effective solid angle of the survey volume
decreases in the same way. However, since the luminosity function is
exponential at the bright end, the magnification significantly
increases the number counts of luminous galaxies
\citep[e.g.][]{2010MNRAS.406.2352L,2013arXiv1301.0360E}. In addition,
the lensing magnification improves the spatial resolution of distant galaxies
\citep[e.g.][]{2008Natur.455..775S,2011MNRAS.410.2506M,2012arXiv1212.6700F,
  2013ApJ...765...48J}. For multiply-imaged systems, the positions
and magnitudes of the images depend on several factors, e.g., lens and
source redshifts, lens mass profiles (including substructures, e.g.
\citealt{1998MNRAS.295..587M}) and intrinsic properties of the background
sources. Thus gravitational lensing provides us with more information for
studying both lens galaxies and background sources.

Lensing magnification itself is achromatic. However, magnification
varies with the position of the source. The spatial profiles at
different wavelengths may not be the same. Thus a source at different
wavelengths may be magnified differently
\citep{1992ARA&A..30..311B}. It has been noticed that differential
magnification can bias the derived results using the spectral energy
distribution (SED) of lensed sources
\citep[e.g.][]{1999MNRAS.304..669B,2001ASPC..237..183P}, or using the
ratio of spectral lines \citep[e.g.][]{1995ApJ...453L..65D}.
\citet{2012ApJ...761...20H} modelled their sources with two components
(a compact and a diffuse one) and showed that the compact one will
usually be magnified by a larger factor than the diffuse one in strong
lensing.

In this paper, we will study the bias in the derived spectral physical
parameters due to differential magnification. We model the background
galaxy as a sum of a bulge and a disk with different sizes
and spectra. Using ray-tracing simulations, the lensing magnifications
of different regions are calculated and applied to obtain the
composite observed spectrum. We start with a discussion of the basic
formalism in section 2, present our model and magnification effects in
section 3, and then discuss our results in section 4.
The cosmology adopted here is a $\Lambda$CDM model with
parameters based on \citet{cfhtls2013kilbinger}:
$\Omega_{\Lambda}=0.718$, $\Omega_{\rm m}=0.286$, $\sigma_8=0.804$,
a Hubble constant $H_0 = 100h$ km\,s$^{-1}$\,Mpc$^{-1}$ with $h=0.693$.

\section{Lensing magnification}

The fundamentals of gravitational lensing can be found in
\citet{2001PhR...340..291B}. The thin-lens approximation is adopted
in this paper, implying that the lens mass distribution can be
projected onto the lens plane perpendicular to the line of sight. We
denote angular coordinates on the lens plane as $\vc\theta$, and those
on the source plane as $\vc\beta$. The lens equation can be written as
\be
\vc{\beta} = \vc{\theta} - {D_{\rm ds} \over D_{\rm s}}\hat\alpha(\vc{\theta})
= \vc{\theta} - \vc{\alpha(\theta)},
\ee
where $D_{\rm ds}$ and $D_{\rm s}$ are the angular diameter distances from
the lens to the source and from the observer to the source respectively.
The deflection angle $\alpha$ can be calculated from the lens model.

We denote the brightness distribution of the source by
$I^s(\vc\beta)$. Since lensing conserves the surface brightness, the
brightness distribution of the image is thus $I(\vc\theta)=
I^s(\vc\beta(\vc\theta))$. The total flux of the source and image are
\be
S_0 = \int \d^2 \beta \; I^s(\vc\beta), \qquad
S= \int \d^2 \theta \; I(\vc\theta),
\elabel{fluximage}
\ee
respectively. The magnification is defined as $\mu\equiv S/S_0$.

\begin{figure}
  \centerline{
    \includegraphics[width=4.8cm,height=4.0cm]{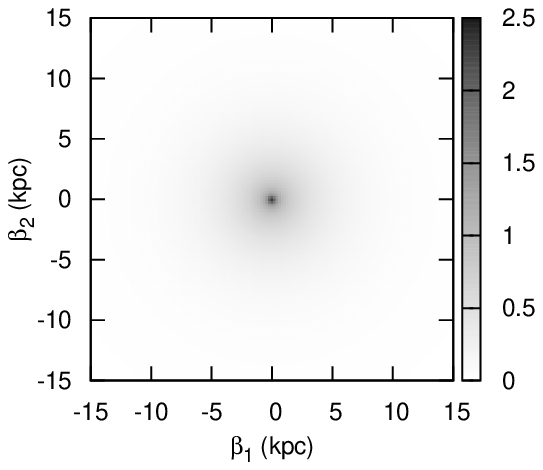}
    \includegraphics[width=4.8cm,height=4.0cm]{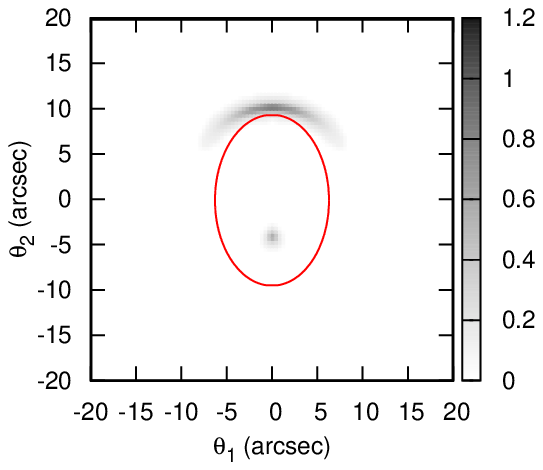}}
  \caption{The simulated image of a background source (left) and the
    lensed (right) images. Notice that the scales and units of the two
    panels are different. In the right panel, the lens is at the
    origin of the map with $\theta_{\rm E}=7.8$ arcsec
    ($\sim15$kpc). The solid line is the critical curve of the
    lens. The lens and source are at redshift 0.1 and 2.0
    respectively.}
  \label{fig:lensimage}
\end{figure}
\begin{figure}
  \centerline{
    \includegraphics[width=4.8cm,height=4.0cm]{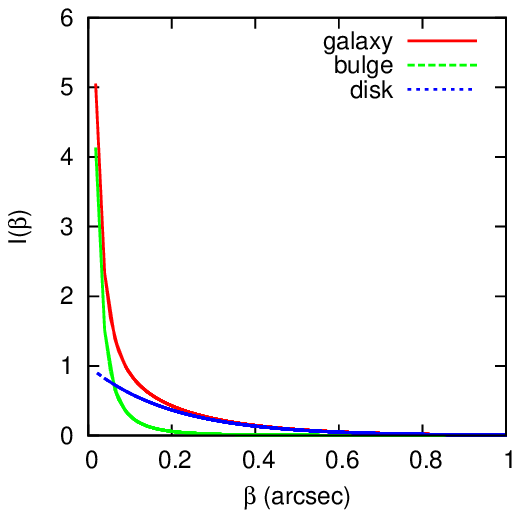}
    \hspace{-0.8cm}
    \includegraphics[width=4.8cm,height=4.0cm]{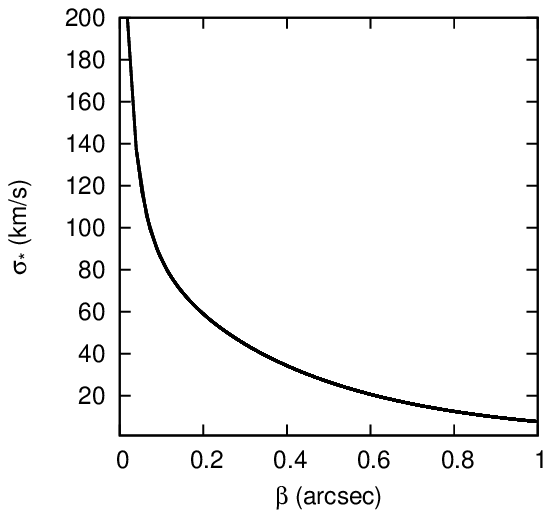}}
  \caption{The left panel shows the surface intensity profile of the
    source galaxy: an axis-symmetric bulge, a face-on disk component
    and the sum. The right panel shows the velocity dispersion
    variation as a function of radius.}
  \label{fig:intensity}
  \vspace{0.2cm}
\end{figure}

\section{Lensing effects on the spectroscopic properties of background sources}

Lensing magnification varies with the relative positions of the lens
and source, especially when the source is close to the caustics. For
extended galaxies, different components differ in intrinsic sizes,
thus their magnifications will be different. Therefore the observed
spectrum of the background galaxy is changed by lensing. The
estimation of galaxy properties differs from its intrinsic ones, like
velocity dispersion ($\sigma_*$), star formation rate (SFR) and
metallicity of galaxies ($Z$). In \citet{2012ApJ...761...20H}, the
galaxy is modelled as the sum of an extended component and an inner
core. The total flux is the sum of the two components and the total
magnification is their weighted average. A similar approach will be
adopted here. In reality, galaxies are structurally complicated,
e.g. they host some clumpy star formation regions. Moreover, their
morphology may not be regular, especially during the early stage of
galaxy formation. The simple model here is for illustration purposes.

\subsection{Light profile of the background galaxy}

The luminosity profiles of the bulge and the disk are described as follows:
\bea
I(R) &=& I_e\, {\rm exp} \{-7.67\eck{\rund{R/R_e}^{1/4}-1}\},
\\
I(R) &=& I_d\, {\rm exp}\rund{-R/R_d},
\elabel{lumprofile}
\eea
where $I_e$ is the surface brightness at the bulge effective radius
$R_e$, $I_d$ is the central surface brightness and $R_d$ is the scale
length of the exponential disk. The bulge to total ratio is
\citep{1987gady.book.....B}
\be
B/T = {R_e^2I_e \over R_e^2I_e +0.28 R_d^2 I_d}.
\ee
We use $I_d=1.0$ and $I_e=0.44$ (the unit is arbitrary here, since we
only need the luminosity ratio between the bulge and the disk). With a small
bulge assumption in this study ($B/T=0.2$)
\citep{2012MNRAS.421.2277L}, we take $R_e=0.7$\,kpc and $R_d=1.7$\,kpc,
corresponding to $0.08$ and $0.2$ arcsec at redshift $z_s=2$ respectively.
For simplicity, we assume an axis-symmetric bulge and a face-on disk
(see the left panel of Fig.~\ref{fig:lensimage}).

For convenience, we will use the angular separation in the rest of
this paper ($1$ arcsec equals to $8.7$ kpc in the source plane
$z_s=2$, and $1.9$\,kpc in the lens plane $z_d=0.1$). The velocity
dispersion is estimated from the surface brightness using the
isothermal sheet. Further approximation, assuming a uniform disc scale
height, the velocity dispersion of the source as a function of radii
is given by $\sigma_*(\theta)\propto \sqrt{I(\theta)}$
\citep[][Chapter~4.4]{1987gady.book.....B}, and we set the central
value as $218\,{\rm km\,s^{-1}}$ (see the right panel of
Fig.~\ref{fig:intensity}).

The stellar velocity dispersion and the metallicity of source galaxy
relies on stellar continuum, while the star formation rate (SFR)
can be estimated from the \oii emission lines. Hence we simulate the
spectra of bulge and disk by combining the emission-line spectra and
the stellar continuum. For stellar continuum, four artificial stellar
populations are used to generate the spectra
\citep{2003MNRAS.344.1000B}. The weight of each stellar population is
set to be equal ($1/4$). Due to the metallicity gradient of galaxies
\citep[e.g.][]{2000A&A...363..537R}, the metallicity of the bulge is
set to be larger than that of the disk region (see Table \ref{tab:ssp}
for more details on the stellar populations of the bulge and disk).

The emission-line spectra of the bulge and disk are also simulated
separately. For the bulge there are two components: a narrow-line
region (NLR) of the central active galactic nucleus (AGN) and a
starburst in the bulge. For the disk there is only a starburst
component.
In order to study the stellar continuum, we can only take into account
the emission of a type II AGN by assuming that the emissions from the
accretion disk and the broad line region are obscured by the torus. The
type II AGN emission-line spectrum is taken from \citet{2008AJ....136.2373R}.
The starburst emission-line spectra are taken from SDSS DR7 with the
selection criteria of [OIII]/$H{\beta} \sim 0.35$ and [NII]/$H{\alpha}
\sim -1.0$, which is appropriate for the star-forming gas rich galaxies at
redshift $2$. The SFRs of the bulge and disk, which are estimated by
using \oii luminosities, are set to be proportional to the surface
brightness.

\begin{deluxetable}{ccc}
\tablecolumns{3}
\tablewidth{0pc}
\tablecaption{Stellar populations used for the bulge and disk}
\tabletypesize{\footnotesize}
\tablehead{
\colhead{Name} & \colhead{Age}   & \colhead{Metallicity $Z/Z_{\odot}$}
}
\startdata
\multicolumn{3}{c}{Bulge} \\ \hline
age116$\_$m42    &   1.0152E+08     &   0.2       \\
age070$\_$m62    &   1.0000E+07     &   1.0       \\
age135$\_$m62    &   9.0479E+08     &   1.0       \\
age150$\_$m72    &   2.5000E+09     &   2.25       \\ \hline
\multicolumn{3}{c}{Disk} \\ \hline
age055$\_$m42    &   5.0100E+06     &   0.2       \\
age070$\_$m42    &   1.0000E+07     &   0.2       \\
age139$\_$m42    &   1.4340E+09     &   0.2       \\
age116$\_$m62    &   1.0152E+08     &   1.0       
\enddata
\label{tab:ssp}
\end{deluxetable}
With the chosen spectra and luminosity profiles of the bulge and the disk,
the lensed spectrum of the background galaxy can be
obtained after knowing the mean magnification of each part of the source:
\be
F(\nu,\vc\theta) = \mu_b F_b(\nu,\vc\theta) + \mu_{d} F_{d}(\nu,\vc\theta).
\ee
where $\mu_b$ and $\mu_d$ are the magnifications of the bulge region
($<0.1$ arcsec) and the disk region, which can be calculated from the
lensing model.

\subsection{Lens modelling}

In order to study the differential magnification effect on the
spectrum of background sources, we perform ray-tracing
simulations. For simplicity, we model our lens as a singular
isothermal ellipsoid with a velocity dispersion of $770$~km/s
($M_{200}\approx 1\times10^{14}\msun$) and ellipticity
$\epsilon=0.2$. The lens is placed at redshift $z_d=0.1$, which gives
an Einstein radius $\theta_{\rm E}=7.8$\, arcsec.
We map a grid of pixels from the lens plane to the source plane using
the lens equation to obtain the surface brightness distribution of the
image. From this distribution, the total flux of the image is obtained
using Eq.~\ref{eq:fluximage}. We will calculate the magnifications
separately for the inner region, $\mu_b$ ($<0.1$ arcsec), and for the
outer region ($\mu_d$). 

\begin{figure}
\centerline{\includegraphics[width=6.0cm,height=5.0cm]{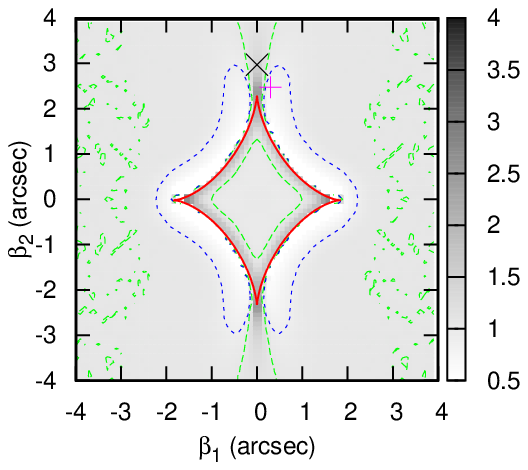}
\hspace{-0.8cm}
\includegraphics[width=6.0cm,height=5.0cm]{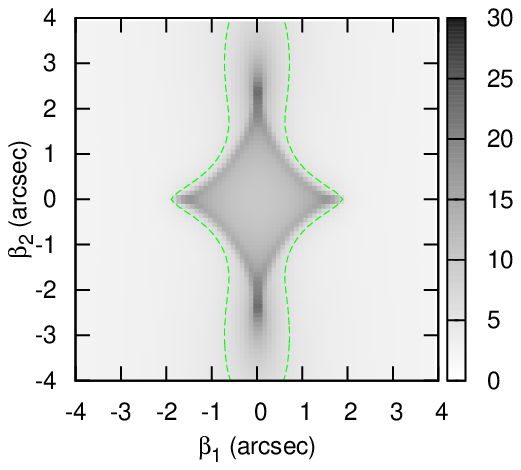}}
\caption{$Left:$ The ratio of $\mu_b/\mu_d$ on the source plane, given
  on linear grey scale and contours. The blue dashed lines enclose the
  region where $\mu_b/\mu_d<0.85$. The green lines enclose the dark
  region where the ratio is greater than $1$. The red solid line is
  the caustics. The cross and plus represent the positions of our
  mock source galaxies. $Right:$ The mean magnification of the whole
  galaxy on the source plane. The green dashed line encloses the region
  where $|\mu|>5$.}
  \label{fig:mubeta}
\end{figure}

The magnification ratio between the bulge and the disk strongly
depends on the relative position of the source and lens, and the
intrinsic size of source galaxies. In the left panel of
Fig.~\ref{fig:mubeta}, one can see that when the bulge of the source
galaxy is close to the caustics, the magnification differences in
the bulge and the disk are significantly larger. An interesting point is
when the disk is closer to the caustics than the bulge, the ratio
becomes smaller than $1$ ($\mu_d>\mu_b$). When the bulge crosses
the caustics, the reverse occurs ($\mu_b>\mu_d$). As a consequence, a
slight shift of the source position (or a change of the lens model) may
cause a significantly different magnification ratio. One can also see
that the region where the differential magnification is strong, the
absolute magnification is also high (e.g. $\mu>5$, right panel of
Fig.~\ref{fig:mubeta}).

We first place the source at $\beta=(0, 3)$ arcsec on the source
plane (and study another position $(0.3, 2.5)$ in the end). Two
images are generated by lensing (see the right panel of
Fig.~\ref{fig:lensimage}). We only study the primary image in this
paper, since it is more luminous and easier to detect. The mean magnification
of the bulge (disk) region is $\mu_b=15$ ($\mu_{d}=10$).

\begin{figure}
\centerline{\includegraphics[width=5.0cm,height=4.2cm]{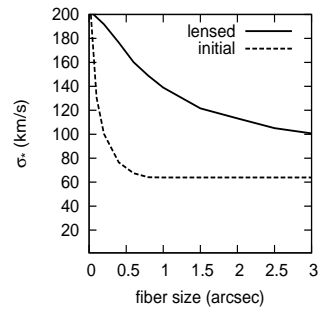}}
\caption{The luminosity-weighted mean velocity
  dispersion as a function of the fiber size.}
\label{fig:obssigma}
\end{figure}

We simulate the velocity dispersion $\sigma_*$ detected by different
sizes of fibers. The centre of the fiber is aligned with the centre of
the lensed source. The mean velocity dispersion is weighted by
luminosity. In Fig.~\ref{fig:obssigma}, the solid and
dashed lines represent the $\sigma_*(\theta)$ after and before lensing
respectively. When the fiber size becomes very small,
the measured velocity dispersion approaches the
intrinsic ones; in the other cases, the measurements of the lensed
galaxy are larger than the initial ones for $\beta=(0, 3)$.
%

\subsection{Results from spectral analysis}
The $\sigma_*$, SFR and metallicity ($Z$) variation of the source and
the lensed galaxy can be obtained through the analysis of the composite
spectra. According to the source model and the parameters listed in
Table \ref{tab:result}, we can fit the galactic continuum using
STARLIGHT \citep{2005MNRAS.358..363C}. This code employs 45 stellar
population templates which are composed of 3 metallicities and 15 ages
\citep{2003MNRAS.344.1000B}. The differences between the spectra of
the lensed image and the initial galaxy are listed in Table
\ref{tab:result}. The velocity dispersion of the lensed image is
higher in the case of $\mu_{b}$/$\mu_{d}=15/10$, since the velocity
dispersion of the bulge is larger than the disk.
In order to get a reliable result, we fit the spectra by using the
direct pixel fitting method
\citep{2006ApJ...641..117G,2009ApJS..183....1H, 2012ApJS..201...31G},
and obtain the same results.

With the continuum subtracted, we can obtain the emission-line only
spectra. The SFR is estimated by fitting [OII]$\lambda$3727. Here the
SFR is normalized to $10 M_{\odot} {\rm yr}^{-1}$
\citep[e.g.][]{2009A&A...504..751S}. The SFR derived from the lensed
image is enhanced by about $11$ times in the case of
$\mu_{b}$/$\mu_{d}=15/10$. After correcting by the mean magnification
of the whole galaxy ($\mu=12.7$), the SFR result is smaller than the
true value by $\sim 10\%$.

The metallicity of the whole galaxy is again obtained from
the STARLIGHT, $Z=\Sigma x_j\times Z_j$, where $x_j$ is the
weight of the $j$-th stellar population, and $Z_j$ is its metallicity.
The result from the lensed image is significantly larger than the initial
one. The reason is that both the metallicity and the magnification of the
bulge are higher than those of the disk, and the weight of the bulge
is enlarged by lensing.

To see the sensitivity of the results to the source position, we place
the source at a different location $(0.3, 2.5)$. The results are given
in the column `lensed 2' in Table~\ref{tab:result}. The magnification
ratio is different from the first case: $\mu_b/\mu_d=7/10$. In this
case, the contribution of the bulge to the total lensed spectrum is
suppressed, which leads to a smaller $\sigma_*$ and metallicity while
the SFR is overestimated.

When we observe the lensed galaxy with fibers, the light may actually
come from different parts of the bulge and the disk, which makes the
measurement of velocity dispersion, SFR and metallicity of the
source more complicated. We perform simulations with a fiber size of
$3$ arcsec diameter. The centre of the lensed image is identified as
the brightest position of the image. The results are given in the
bottom part of Table~\ref{tab:result}. For both source positions,
similar biases are obtained.


\begin{deluxetable*}{cccccc}
\tablecolumns{5}
\tablewidth{0pc}
\tablecaption{Results of differential magnification}
\tabletypesize{\footnotesize}
\tablehead{
\colhead{Parameters} &\colhead{Bulge}  &\colhead{Disk}
&\colhead{lensed $1$}  &\colhead{lensed $2$}  &\colhead{Source}}
\startdata
\multicolumn{6}{c}{total flux} \\ \hline
$\mu$                    &  15(7)    &  10(10)   & 15+10    & 7+10   & 1+1   \\
luminosity                   &  1        &  4        & 58       & 50     & 5 \\
$\sigma_{*}$(km/s)   &  131      &   48      &    67    &  58    & 59    \\
SFR (10$\msun$/yr)            &    2      &    8      & 113(8.9)    &93(11)   & 10    \\
metallicity $Z/Z_\odot$      &   0.95    &   0.4     &   0.59   & 0.47   & 0.51   \\ \hline
\multicolumn{6}{c}{flux within 3 arcsec diameter fiber} \\ \hline
$\mu$                    &  15(6.7)  &13.5(6.6)  & 15+13.5  &6.7+6.6 & 1+1   \\
luminosity                   &  1        &  0.67     &  19      & 11     & 1.67 \\
$\sigma_{*}$(km/s)   &  164(134) &  53(52)   &  118     & 76     & 110 (74)    \\
SFR (10$\msun$/yr)            &    6      &    4      & 144(9.6)   & 67(10)   & 10    \\
metallicity $Z/Z_\odot$      &    0.95   &   0.4     &   0.8    & 0.58   & 0.61   \\
\enddata
\tablecomments{ The column `lensed 1' represents the lensed results
  for the source position $(0, 3)$ (black cross in
  Fig.~\ref{fig:mubeta}), while the column `lensed 2' represents the
  results for the source position $(0.3, 2.5)$ (purple plus in
  Fig.~\ref{fig:mubeta}).  $\sigma_{*}$ is the velocity dispersion
  estimated using the method in \citet{2012ApJS..201...31G}. The SFR
  in brackets is corrected by the mean magnification of the whole
  lensed galaxy.}
\label{tab:result}
\end{deluxetable*}

\section{Summary and discussion}

In this paper, we have studied the differential magnification effect
on the spectrum of a lensed background galaxy. Two components (a bulge
and a disk) with different luminosity profiles and spectra are used to
model the background galaxy. Ray-tracing simulations are employed to
calculate the lensing magnification. We find that the derived
properties of the lensed galaxy are changed, e.g. SFR, metallicity and
velocity dispersion. Velocity dispersion and metallicity can either
increase or decrease, depending on the relative position of the source
and the lens. The SFR is strongly enhanced but can be corrected by
dividing the mean magnification. However, a slight bias of SFR still
remains after correction.
Not surprisingly, the differential magnification will be strong
when the source is close to the caustics, and the absolute value of the
magnification is also high (e.g. $|\mu|>5$). In this case, a slight shift
of the source position will cause a dramatic change in the derived
results. Therefore, one needs to take into account the differential
magnification effect for highly magnified extended sources.

Because of observational limitations, the black hole vs. velocity
dispersion ($M_{\rm BH}-\sigma_*$) relation at high redshift
\citep[e.g.][]{2002ApJ...574..740T} is uncertain. Gravitational
lensing boosts our chances of studying black holes at high
redshift. However, we found that the measured $\sigma_*$ may be an
unreliable indicator of the black hole mass.
In practice, for very high-redshift galaxies, it is easier to observe
with slits than fibres since the background subtraction is more
reliable. To avoid extra bias entering the results, one needs to
consider carefully the region of the background galaxy covered when
taking a spectrum.

Our study is simplistic in several ways. For example, source galaxies may be
irregular, lacking regular bulge and disk components. Additionally,
high redshift galaxies can be complicated, e.g. they may have
clumpy star forming regions with different spectra
\citep{2008ApJ...687...59G,2013ApJ...765...48J}. This will mislead
the calculation of derived physical properties. Moreover, lens clusters are also
complicated. The magnification map is strongly affected by the shape
and mass profile (including substructures) of the lens. Therefore
accurate lens modelling is necessary when deriving the properties of
lensed galaxies.

On the other hand, multiple images can provide more constraints on
both the lens and source. With Integral Field Unit (IFU) spectra for
different components of the multiple images at high redshift, we can study
the lens mass distribution and the intrinsic properties of background
galaxies in greater detail. With iterative modelling of the lens and
source, a relatively precise source image can be reconstructed. The
bias due to differential magnification can be strongly suppressed.

Current telescopes, such as the Hubble Space Telescope (HST) or the
Keck telescope can already study differential magnifications. The
next-generation telescopes with adaptive optics, such as the Thirty
Meter Telescope and the E-ELT, will allow even more
accurate studies of dynamics in high redshift galaxies.

\section{Acknowledgments}
We thank Richard Long, Tucker Jones, and the referee for useful comments on the
draft. XE is supported by NSFC grant No.11203029. SM is supported by
the Chinese Academy of Sciences and the National Astronomical
Observatories of China.
\bibliographystyle{apj}
\bibliography{../../../bib/lens,../../../bib/smg,../../../bib/refcos,../../../bib/stronglens,../../../bib/galaxy,../../../bib/refbooks,../../../bib/qso}

\end{document}